# PepSIRF + QIIME 2: software tools for automated, reproducible analysis of highly-multiplexed serology data

Annabelle M. Brown[1], Evan Bolyen[1], Isaiah Raspet[1], John A. Altin[2] and Jason T. Ladner[1]*

[1]The Pathogen and Microbiome Institute, Northern Arizona University, Flagstaff, AZ 86011
[2]TGen, Flagstaff, AZ 86005

*Corresponding author: jason.ladner@nau.edu

PepSIRF (Fink et al. 2020) is a command-line, module-based open-source software package that facilitates the analysis of data from highly-multiplexed serology assays (e.g., PepSeq or PhIP-Seq). It has nine separate modules in its current release (v1.5.0): *demux*, *info*, *subjoin*, *norm*, *bin*, *zscore*, *enrich*, *link*, and *deconv*. These modules can be used together to conduct analyses ranging from demultiplexing raw high-throughput sequencing data to the identification of enriched peptides. QIIME 2 (Bolyen et al. 2019) is an open-source, community-developed and plugin-based bioinformatics platform that focuses on data and analytical transparency. QIIME 2's features include integrated and automatic tracking of data provenance, a semantic type system, and built-in support for many types of user interfaces. Here, we describe three new QIIME 2 plugins that allow users to conduct PepSIRF analyses within the QIIME 2 environment and extend the core functionality of PepSIRF in two key ways: 1) enabling generation of interactive visualizations and 2) enabling automation of analysis pipelines that include multiple PepSIRF modules.

The three plugins (*q2-pepsirf*, *q2-ps-plot*, and *q2-autopepsirf*) are open-source, multiple interface programs written in Python 3 and run within the QIIME 2 Framework. The *q2-pepsirf* plugin is a QIIME 2 wrapper of the core functionality of PepSIRF, and defines all of the file formats for the remaining plugins. The *q2-ps-plot* plugin is a *q2-pepsirf* dependent plugin that is used for visualizing the results of PepSIRF analyses. Lastly, *q2-autopepsirf* is dependent on both *q2-pepsirf* and *q2-ps-plot*, and it provides pipelines that automate the process of running *q2-pepsirf* and *q2-ps-plot* for several common analysis workflows. **Figure 1** depicts the flow of the dependencies of the three plugins.

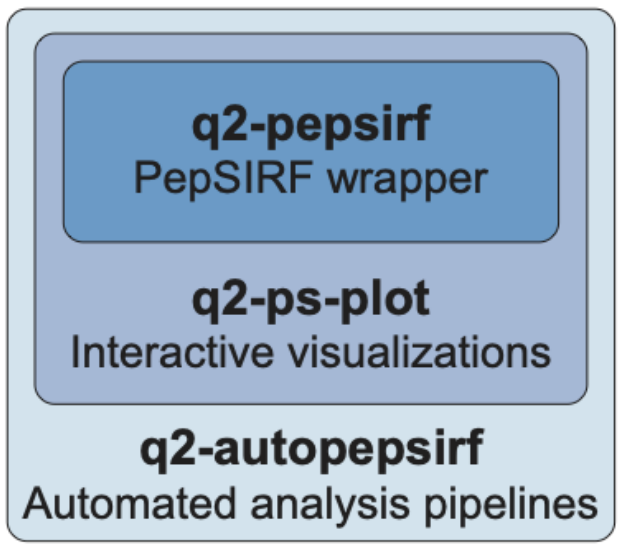

**Figure 1.** A graphical depiction of the flow of dependencies of the three plugins. The *q2-pepsirf* plugin is dependent only on itself, *q2-ps-plot* is dependent on *q2-pepsirf*, and *q2-autopepsirf* is dependent on both *q2-pepsirf* and *q2-ps-plot*.

By implementing (and expanding) the functionality of PepSIRF within QIIME 2 plugins, we are able to take advantage of the features of QIIME 2 that promote transparency and repeatability. Data provenance tracking allows users to view how a QIIME 2 result was generated. This will include information on all QIIME 2 actions that were run, including all parameter values and universally unique identifier (UUID) references to all inputs and outputs (Bolyen et al. 2019). QIIME 2 also offers a semantic type system, which prevents common analysis errors by reducing the allowable inputs of a step to only those which make semantic sense and provides a common ground for community development and extension as it provides a shared vocabulary for data. Lastly, there is support for multiple user interfaces, such as the command-line, Python 3 Application Programming Interface (API), and recently, Galaxy (Afgan et al. 2018). This allows users to choose which interface best fits their needs, such as a more advanced/technical user taking advantage of the Python 3 API, which allows the user to directly interact with the plugins from within a Python 3 script.

## q2-pepsirf: a QIIME 2 wrapper for PepSIRF

The *q2-pepsirf* plugin (https://github.com/LadnerLab/q2-pepsirf) serves as a wrapper for the PepSIRF binary and runs PepSIRF within QIIME 2. This plugin supports running all nine PepSIRF modules, but these have been implemented as 11 different actions within the plugin in order to handle different input file types and optional outputs: *demux*, *infoSNPN*, *infoSumOfProbes*, *subjoin*, *norm*, *bin*, *zscore*, *enrich*, *link*, *deconv-singular*, and *deconv-batch*. This plugin allows users to access the full functionality of PepSIRF from within the QIIME 2 environment and facilitates the generation of custom workflows. It also serves as a dependency for the *q2-ps-plot* and *q2-autopepsirf* plugins. The output from each action will be one or more QIIME 2 artifact (QZA) files that can be used for further analysis within QIIME 2 or can be exported into a user readable file type, such as a tab-separated value (TSV) file. Usage information and tutorials are available here: https://ladnerlab.github.io/pepsirf-q2-plugin-docs/.

## q2-ps-plot: visualizing highly-multiplexed serology data

### General Design

The *q2-ps-plot* plugin (https://github.com/LadnerLab/q2-ps-plot) utilizes the outputs from PepSIRF (along with additional metadata) to generate a variety of interactive visualizations that facilitate quality control checks and the interpretation of results. These visualizations are generated using Altair (https://altair-viz.github.io/). The current version of *q2-ps-plot* (v2021.11.12+73.g38779ed) features six visualizers and four TSV/Dir pipelines (see below for details). The visualizers require QZA files (such as those generated by *q2-pepsirf*) as inputs. However, the TSV/Dir pipelines (https://ladnerlab.github.io/pepsirf-q2-plugin-docs/tsv-pipelines/) allow the user to provide alternative file types (e.g., TSV, TXT) as input or directories of such files, as appropriate. Each of the visualizers and pipelines of this plugin will output a QIIME 2 visualization (QZV) file that is viewable with QIIME 2 View (https://view.qiime2.org/) and includes the data provenance information. The visualizers/pipelines available within this plugin are:

*repScatters/repScatters-tsv*, *zenrich/zenrich-tsv*, *proteinHeatmap/proteinHeatmap-dir*, *mutantScatters/mutantScatters-tsv*, *enrichmentRCBoxplot*, and *readCountsBoxplot*. Usage information and tutorials are available here: https://ladnerlab.github.io/pepsirf-q2-plugin-docs/.

## Visualizers and Pipelines

### repScatters and repScatters-tsv

The **repScatters** visualizer and **repScatters-tsv** pipeline facilitate visual comparisons between assays using peptide-level normalized read counts or enrichment Z scores. This is a critical component of quality control when running multiple technical replicates and can also be used to compare overall patterns of reactivity between distinct biological samples. In **repScatters**, the user must provide a "pairs source file" and either a Z score QZA file (FeatureTable[Zscore], e.g., the output from the *zscore* action of *q2-pepsirf*) or a normalized read count QZA (FeatureTable[Normed], e.g., the output from the *norm* action of *q2-pepsirf*). The pairs source file should be tab-delimited and contain information for sample replicates (one line per pair of samples/replicates to compare). The QZV output will contain an interactive visualization (**Figure 2**) with scatter plot heatmaps for each specified pair of samples. The user can toggle between samples using a dropdown menu. The **repScatters-tsv** pipeline allows for a repScatters analysis with TSV (rather than QZA) inputs.

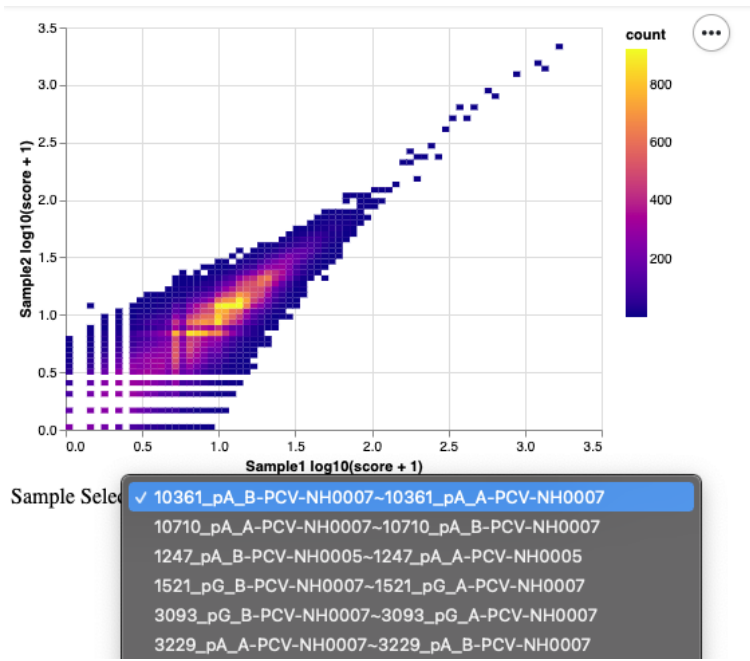

**Figure 2.** An example QZV output from **repScatters** viewed using QIIME 2 View. This screenshot was taken while engaging with the interactive sample selection dropdown menu.

**zenrich and zenrich-tsv**

The *zenrich* visualizer and *zenrich-tsv* pipeline are used to create scatter plot heatmaps that compare normalized read counts between an experimental sample (y-axis) and relevant negative controls (x-axis), while also highlighting putatively enriched peptides according to user-defined Z score thresholds. The points of the scatter plot are colored according to the Z score thresholds provided, and the scatter plot is layered over a heatmap containing all of the data for the sample selected. For the *zenrich* visualizer, QZA files are expected as inputs, while the *zenrich-tsv* pipeline supports the use of TSV files as input. There are a number of user-defined thresholds and optional parameters that allow for a high degree of customization. The QZV output will contain an interactive visualization (**Figure 3**) with a dropdown menu for the user to specify which sample to view, as well as a tooltip that displays information about specific enriched peptides based on input from the user (i.e., when the cursor is placed above the corresponding point).

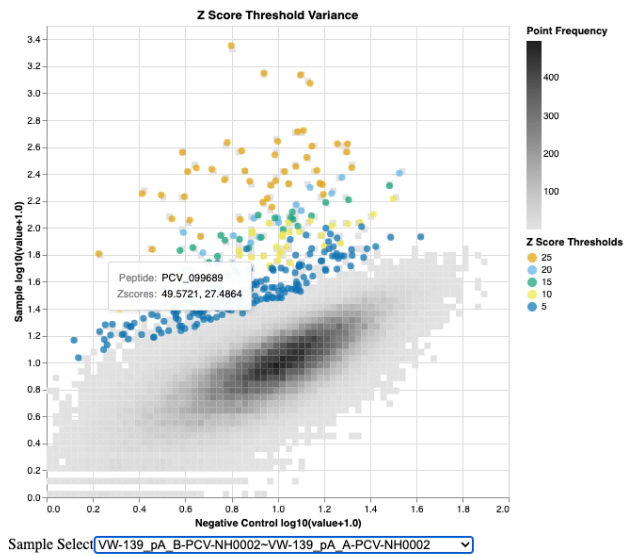

**Figure 3.** An example QZV output from *zenrich* viewed using QIIME 2 View. This screenshot depicts the interactive tooltip that is available for obtaining more information about the enriched peptides (colored circles).

**proteinHeatmap and proteinHeatmap-dir**

The *proteinHeatmap* visualizer and *proteinHeatmap-dir* pipeline generate heatmaps of enriched peptides based on the positions of the peptides within the proteins from which they were designed. Two main user inputs are required: 1) a directory containing ≥1 enriched peptide files (one per sample, e.g., the output of the *enrich* action of *q2-pepsirf*) and 2) protein alignment files containing alignment coordinates for peptides within each protein of interest. For the *proteinHeatmap* visualizer, inputs must be provided in QZA format, while for the *proteinHeatmap-dir* pipeline, QZA formatted inputs are not required (i.e., a simple directory path can be provided for the enriched peptide files and TSV formated files can be provided for the protein alignment coordinates). The QZV output will contain an interactive heatmap (**Figure 4**) with plots for each protein, each of which can be selected using the dropdown menu, and a tooltip that highlights one of the enriched peptides at each position in the alignment (Note: only one peptide name is displayed even if multiple peptides overlap the indicated position).

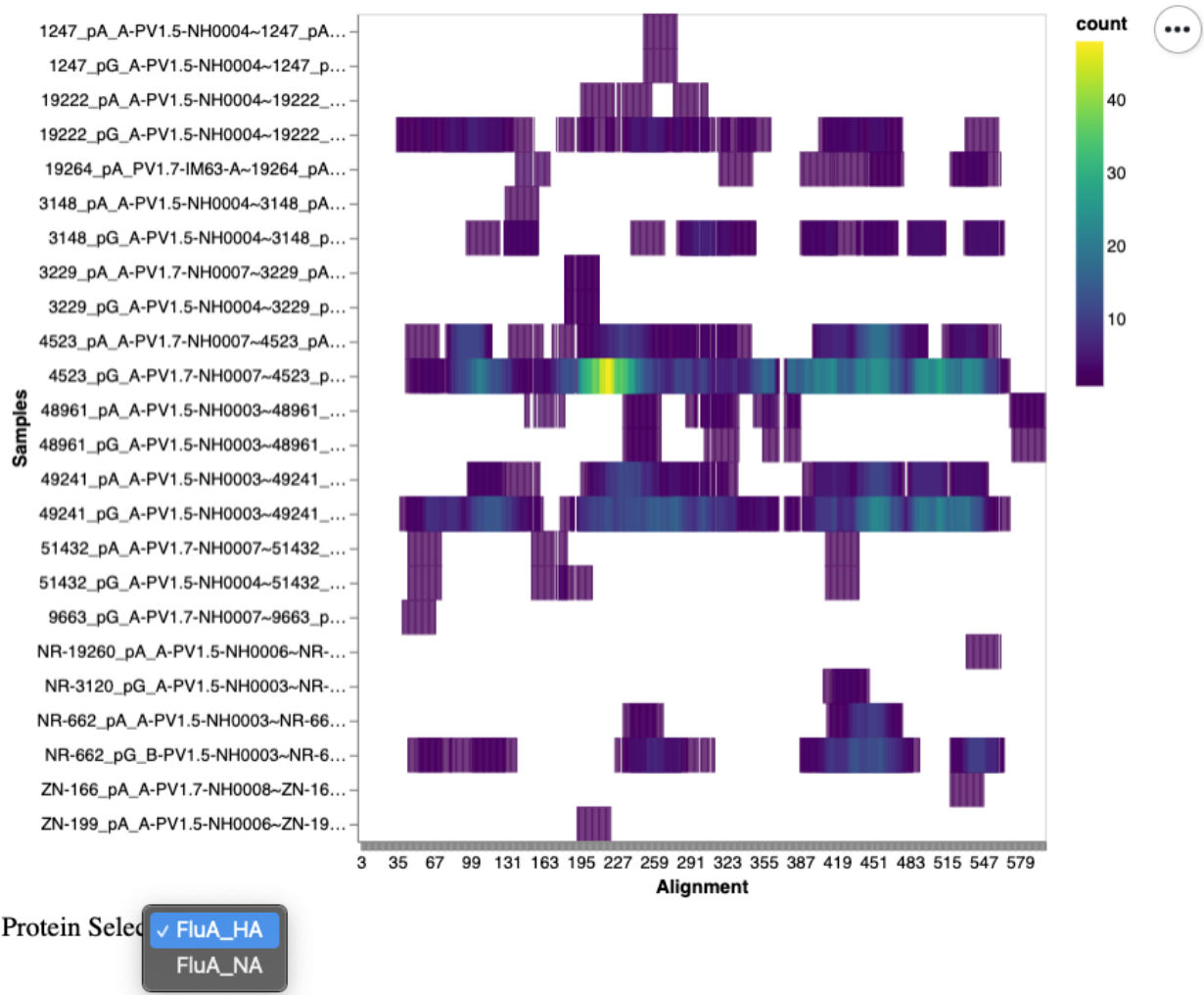

**Figure 4.** An example QZV output from *proteinHeatmap* viewed using QIIME 2 View. This screenshot was taken while engaging with the interactive sample selection dropdown menu.

**mutantScatters and mutantScatters-tsv**

      The **mutantScatters** visualizer and **mutantScatters-tsv** pipeline are used to generate a scatterplot (and optional boxplot) comparing reactivity against sets of related peptides contained in the same assay (e.g., alanine scanning or site saturation mutagenesis). A reference (or wildtype) peptide within the design should be indicated for each set of related peptides, and a single interactive visualization can include data from multiple samples and from several distinct sets of peptides. The QZV output will contain an interactive visualization (**Figure 5**) with plots for each sample/peptide set combination. The user can toggle between samples and peptide sets using the dropdown menus, scatter points can be colored according to a user-defined category, and there is an interactive tooltip that can display information associated with a peptide of interest. The **mutantScatters** visualizer requires QZA formatted input files, while the **mutantScatters-tsv** pipeline requires TSV formatted input files.

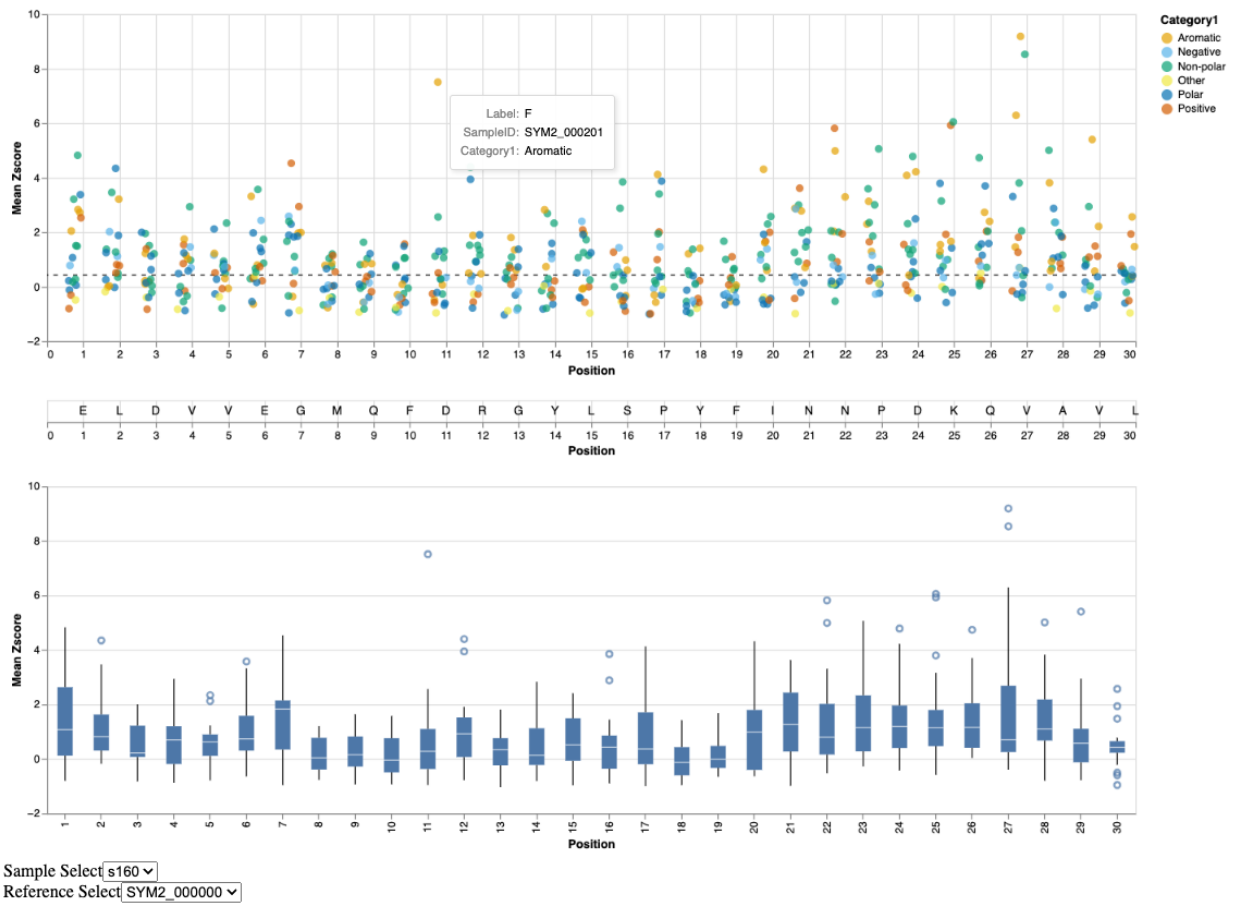

**Figure 5.** An example QZV output from **mutantScatters** viewed using QIIME 2 View. This screenshot depicts the interactive tooltip that is available for obtaining more information about the individual peptides.

**readCountsBoxplot and enrichmentRCBoxplot**

The *readCountsBoxplot* and *enrichmentRCBoxplot* visualizers are used to generate boxplots summarizing total raw read counts or number of enriched peptides, respectively, across all samples in an analysis. These visualizers are most useful when integrated within an automated pipeline, e.g., the *diffEnrich* pipeline within the *q2-autopepsirf* plugin. Each of these visualizers generates two outputs: 1) an interactive QZV file viewable with QIIME 2 View (**Figure 6**) and 2) a static PNG file. The QZV output includes a hoverable tooltip that displays the associated summary statistics (minimum, maximum, median, etc.).

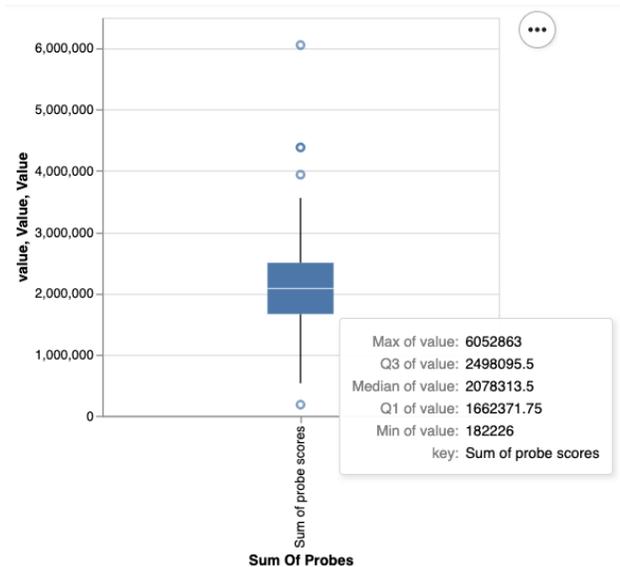

**Figure 6.** An example QZV output from *readCountsBoxplot* viewed using QIIME 2 View. This screenshot depicts the interactive tooltip that is available for viewing summary statistics.

# q2-autopepsirf: automated analysis and visualization pipelines

## General Design

The *q2-autopepsirf* plugin (https://github.com/LadnerLab/q2-autopepsirf) automates commonly used data analysis pipelines that each involve multiple actions and visualizers from the *q2-pepsirf* and *q2-ps-plot* plugins. The pipelines within *q2-autopepsirf* generate all of the QZA and QZV files that would be generated from running each component action individually, and the user can also optionally generate corresponding TSV outputs for each QZA. Built-in QIIME 2 provenance tracking is also available for these pipelines, however, currently, the provenance is only viewable by accessing individual components of a given QZA file, which is a limitation of the QIIME 2 View interface. In the future, we hope to make this provenance more easily accessible. The current version of *q2-autopepsirf* (v2021.12+37.g7b5cd40) features two QZA pipelines and two corresponding TSV pipelines. Usage information and tutorials are available here: https://ladnerlab.github.io/pepsirf-q2-plugin-docs/.

## Pipelines

**diffEnrich and diffEnrich-tsv**

The ***diffEnrich*** and ***diffEnrich-tsv*** actions automate the analysis of highly-multiplexed serology data starting with raw read counts and ending with lists of enriched peptides for each sample. This pipeline has been described in multiple publications (Ladner et al. 2021; Elko et al. 2022) and includes use of the *norm*, *zscore,* and *enrich* actions of *q2-pepsirf* to normalize the data (for differences in sequencing depth and relative to negative controls), calculate Z scores, and generate lists of enriched peptides. These pipelines also automate the running of several *q2-ps-plot* visualizers (*enrichmentRCBoxplot*, *readCountsBoxplot*, *zenrich*, *repScatters* for Z scores and *repScatters* for normalized read counts) to aid in quality control assessment. The ***diffEnrich*** pipeline requires QZA formatted input files, while the ***diffEnrich-tsv*** pipeline requires TSV formatted input files.

**diffEnrichDeconv and diffEnrichDeconv-tsv**

The ***diffEnrich-deconv*** and ***diffEnrich-deconv-tsv*** pipelines automate the analysis of highly-multiplexed serology data starting with raw read counts and ending with lists of enriched peptides for each sample and lists of viruses to which the individual is predicted to have previously been exposed. These pipelines are equivalent to ***diffEnrich*** and ***diffEnrich-tsv***, respectively; however, the resulting lists of enriched peptides are then run through the *deconv-batch* action of *q2-pepsirf* to deconvolute the list of peptides into predicted seropositivities.

## Acknowledgements

Development of the software reported in this publication was supported by the National Institute of Allergy and Infectious Diseases of the National Institutes of Health under award number U24AI152172 (PI: John Altin) and contract number W15QKN-20-9-C003 (PI: Paul Keim) in support of the Defense Threat Reduction Agency. IR was also supported by the National Science Foundation through the Louis Stokes Alliances for Minority Participation (LSAMP) program. The content is solely the responsibility of the authors and does not necessarily represent the official views of the U.S Government.

## References

Afgan, Enis, Dannon Baker, Bérénice Batut, Marius van den Beek, Dave Bouvier, Martin Cech, John Chilton, et al. 2018. "The Galaxy Platform for Accessible, Reproducible and Collaborative Biomedical Analyses: 2018 Update." *Nucleic Acids Research* 46 (W1): W537–44.
Bolyen, Evan, Jai Ram Rideout, Matthew R. Dillon, Nicholas A. Bokulich, Christian C. Abnet, Gabriel A. Al-Ghalith, Harriet Alexander, et al. 2019. "Reproducible, Interactive, Scalable and Extensible Microbiome Data Science Using QIIME 2." *Nature Biotechnology* 37 (8):

852–57.

Elko, Evan A., Georgia A. Nelson, Heather L. Mead, Erin J. Kelley, Virginia Le Verche, Angelo A. Cardoso, Jennifer L. Ely, et al. 2022. "COVID-19 Vaccination Recruits and Matures Cross-Reactive Antibodies to Conserved Epitopes in Endemic Coronavirus Spike Proteins." *medRxiv : The Preprint Server for Health Sciences*, January. https://doi.org/10.1101/2022.01.24.22269542.

Fink, Zane W., Vidal Martinez, John Altin, and Jason T. Ladner. 2020. "PepSIRF: A Flexible and Comprehensive Tool for the Analysis of Data from Highly-Multiplexed DNA-Barcoded Peptide Assays." *arXiv Preprint arXiv:2007. 05050*.

Ladner, Jason T., Sierra N. Henson, Annalee S. Boyle, Anna L. Engelbrektson, Zane W. Fink, Fatima Rahee, Jonathan D'ambrozio, et al. 2021. "Epitope-Resolved Profiling of the SARS-CoV-2 Antibody Response Identifies Cross-Reactivity with Endemic Human Coronaviruses." *Cell Reports. Medicine* 2 (1): 100189.